\documentclass{aastex631}
\usepackage[utf8]{inputenc}
\usepackage{booktabs}
\usepackage{color,comment}

\begin{document}
\title{Tracing the History of Obscured Star Formation with Cosmological Galaxy Evolution Simulations}
\author[0009-0008-7017-5742]{Dhruv T. Zimmerman}
\affiliation{Department of Astronomy, University of Florida, 211 Bryant Space Sciences Center, Gainesville, FL 32611 USA}

\author[0000-0002-7064-4309]{Desika Narayanan}
\affiliation{Department of Astronomy, University of Florida, 211 Bryant Space Sciences Center, Gainesville, FL 32611 USA}
\affiliation{University of Florida Informatics Institute, 432 Newell Drive, CISE Bldg E251, Gainesville, FL 32611}
\affiliation{Cosmic Dawn Center (DAWN), Niels Bohr Institute, University of Copenhagen, Jagtvej 128, København N, DK-2200, Denmark}

\author[0000-0001-7160-3632]{Katherine E. Whitaker}
\affiliation{Department of Astronomy, University of Massachusetts, Amherst, MA 01003, USA}
\affiliation{Cosmic Dawn Center (DAWN), Niels Bohr Institute, University of Copenhagen, Jagtvej 128, København N, DK-2200, Denmark}

\author[0000-0003-2842-9434]{Romeel Dav\'e}
\affiliation{SUPA, Institute for Astronomy, Royal Observatory, Blackford Hill, Edinburgh, EH9 3HJ, UK}
\affiliation{University of the Western Cape, Bellville, Cape Town 7535, South Africa}
\affiliation{South African Astronomical Observatories, Observatory, Cape Town 7925, South Africa}

\date{December 2023}

\begin{abstract}
    We explore the cosmic evolution of the fraction of dust obscured star formation predicted by the \textsc{simba} cosmological hydrodynamic simulations featuring an on-the-fly model for dust formation, evolution, and destruction. We find that up to $z=2$, our results are broadly consistent with previous observational results of little to no evolution in obscured star formation. However, at $z>2$ we find strong evolution at fixed galaxy stellar mass towards greater amounts of obscured star formation. We explain the trend of increasing obscuration at higher redshifts by greater typical dust column densities along the line of sight to young stars. We additionally see that at a fixed redshift, more massive galaxies have a higher fraction of their star formation obscured, which is explained by increased dust mass fractions at higher stellar masses. Finally, we estimate the contribution of dust-obscured star formation to the total star formation rate budget and find that the dust obscured star formation history (SFH) peaks around $z\sim 2-3$, and becomes subdominant at $z\ga 5$. 
\end{abstract}

\section{Introduction}\label{sec:intro}
Intimately tied with an understanding of the evolution of our universe is the history of cosmic star formation. Star formation is responsible for many key processes that govern how galaxies evolve with time, such as depletion of galactic gas, stellar feedback into the interstellar medium (ISM), and chemical enrichment of the ISM for future generations of stars (see reviews such as \citealt{elmegreen_turb1_2004,scalo_turb2_2004,veilleux_wind_2005,peroux_cycle_2020,tacconi_ism_2020}). Current observational constraints suggest that the majority of the star formation in the universe took place during the epoch often referred to as `cosmic noon', at $z\sim1-3$ (e.g. \citealt{madau_cosmic_2014}), though much of this star formation may be enshrouded in dust \citep{lefloch_spitzer_2009,bouwens_alma_2016,dunlop_deep_2017,laporte_dust_2017,casey_mapping_2021}.
Therefore, in order to understand galaxy formation and evolution, we need to also understand the evolution in the amount of dust obscured star formation at low and high redshift across the stellar mass function.

The current consensus from observations is that the relative amount of dust-obscured star formation at a fixed galaxy stellar mass does not strongly evolve with time up to $z\sim3$ (e.g. \citealt{bouwens_alma_2016,whitaker_constant_2017,bourne_evolution_2017,mclure_dust_2018}) or roughly 85\% of cosmic history. More recent higher redshift observations suggest that galaxies beyond $z\gtrsim4$ may have less of their star formation obscured (e.g., \citealt{fudamoto_alpine_2020,algera_dustobscured_2023}).
Most of these results suggest that the main dependence of dust obscuration is on a galaxy's stellar mass; more massive galaxies are typically more obscured.
\cite{bouwens_alma_2016} and \cite{whitaker_constant_2017} explore dust-obscured star formation by comparing the relative amount of light emitted from galaxies in samples stacked by stellar mass at the rest-frame ultraviolet (UV) and infrared (IR) wavelengths and see little redshift evolution. 
\cite{mclure_dust_2018} and \cite{shapley_mosfire_2022} instead investigate dust attenuation curves up to redshifts of $\sim3$ and similarly find minimal evolution in the inferred dust-obscured star formation.
At the same time,  investigations of dust-obscured star formation leveraging ALMA at higher redshifts ($z\sim4-6$) suggest that galaxies instead exhibit a lower fraction of obscured star formation relative to galaxies today \citep[e.g.][]{fudamoto_alpine_2020,gruppioni_alpine-alma_2020,algera_dustobscured_2023}.
Taken together, observations suggest an epoch of $z \sim 4$ as the transition period after which
dust-obscured star formation becomes the dominant contributor to the total star formation budget \citep{casey_mapping_2021,zavala_evolution_2021}.

Observations from low and high redshift suggest a relatively consistent picture of dust-obscured star formation. At low redshift, a greater fraction of star formation is obscured by dust than for high-redshift galaxies. However, the nature of determining the fraction of dust obscured star formation observationally is subject to significant uncertainties owing to selection effects; UV/optically selected samples will preferentially be unobscured, and the opposite is true for IR selected samples. Similarly, different observational tracers for the star formation rate in galaxies can result in systematically different inferred values of dust obscured star formation \citep[e.g.][]{kennicutt12a}. As a result, simulations may help to elucidate the situation by virtue of knowledge of the ground truth. 

Theoretical work in this area has focused on particular aspects of obscured star formation, and it is often restricted to high redshift. Theoretical have explored cosmological simulations \citep{ma_fire2lum_2018,vogelsberger_jwsttng_2020,shen_jwsttng_2022,lewis_dustier_2023}, zoom-in simulations \citep{pallottini_serra_2022,vijayan_flares_2023}, and semi-analytic models \citep{mauerhofer_delphi_2023}. However, each of these is limited in at least one key aspect necessary to study obscured star formation across cosmic time including: (1) the dust is post-processed onto the simulation and therefore not produced in a self-consistent manner, (2) the sample of galaxies is limited, or (3) the results do not extend to low redshift comparisons and cannot address the question of how dust-obscured star formation might change with cosmic time. We attempt to address all of these limitations in this work.

In this paper, we generate and analyze synthetic observations of galaxies with dust utilizing the \textsc{simba} cosmological simulation to investigate dust-obscured star formation across both the galaxy mass function, and cosmic time. We focus on whether we can reproduce observational results and then leverage our simulations to understand the physics driving our results. In \S~\ref{sec:meth}, we present our methodology for producing a mock galaxy SED sample; in \S~\ref{sec:res}, we present our predictions for dust-obscured star formation in our simulation up to $z=6$ and use our knowledge of the simulation physics to investigate its causes. In \S~\ref{sec:dis}, we provide discussion of our results in context of the literature, and in \S~\ref{sec:sum} we summarize our work.

\section{Methodology}\label{sec:meth}
\subsection{Simulations}\label{subsec:simba}

In order to understand the impact of dust obscuration on our knowledge of star formation, we utilize the \textsc{simba}
cosmological simulations \citep{dave_simba_2019} and construct synthetic observations of identified galaxies within \textsc{simba}. Here, we include a summary of the relevant physics in \textsc{simba}. For a full discussion, see \cite{dave_simba_2019}. 

\textsc{simba} is built off of the \textsc{gizmo} hydrodynamic code \citep{hopkins_new_2015} and is the successor to the \textsc{mufasa} simulations \citep{dave_mufasa_2016}. It features a star formation rate (SFR) dependent on the $\text{H}_2$ density \citep{kennicutt_global_1998} and follows a Chabrier initial mass function (IMF) for its stellar mass loss \citep{chabrier_imf_2003}. The $\text{H}_2$ mass is determined based on the \citet{krumholz_h2_2011} model. The H gas cooling is calculated by the \textsc{GRACKLE} chemistry module \citep{smith_grackle_2017} with the self-shielding model of \cite{rahmati_hicolumn_2013} against an ionizing background calculated from the \cite{haardt_radiative_2012} model. Stellar feedback is present in metal-loaded winds that enrich the ISM for future generations of stars. Additionally, \textsc{simba} includes black hole growth based on two growth modes: a cold accretion mode based on \cite{anglesalcazar_torque_2017} and a hot halo mode based on \cite{bondi_accretion_1944}. Black hole feedback takes both kinetic and radiative forms; kinetic feedback is jet-based for low accretion and wind-based for high accretion, while radiative feedback is based off of an X-ray model dumping energy into gas. The black hole model for X-ray feedback is derived from \cite{choi_xray_2012}.

Importantly, \textsc{simba} features a self-consistent dust production, growth, and destruction model \citep{dave_simba_2019,li_dust--gas_2019}. {\sc simba} tracks the presence of 11 elements across cosmic time produced by an enrichment model based on Type II and Type 1a supernovae and Asymptotic Giant Branch (AGB) stars. All dust grains are assumed to have the same constant size of 0.1 $\mu$m and have their motions tied to the gas particles. 
Dust production is governed by taking fixed fractions of these metals from the enrichment process of Type II supernovae and AGB star production as in the \cite{dwek_abundance_1998}
model, with newer models for the condensation based on the models of \cite{ferrarotti_dustcomp_2006} and \cite{bianchi_dustform_2007}. The grains grow by accreting local gaseous metals. Dust is destroyed by thermal sputtering and a subgrid model for supernova shocks \citep{mckinnon_dustform_2016}. Gas particles can also have their dust destroyed by the models for AGN jet feedback \citep{anglesalcazar_torque_2017}, AGN X-ray emission \citep{choi_xray_2012}, winds, and star formation. This model is tuned to reproduce the $z=0$ dust mass function and is successful in reproducing higher redshift dust measurements \citep{li_dust--gas_2019}.

We analyze the high-resolution \textsc{simba} m25n512 run - this is a simulated cosmological box with $25/h$ Mpc sides in comoving coordinates that contains $512^3$ particles and a mass resolution of $2.3\times10^6 M_{\odot}$ and $1.2\times10^7 M_{\odot}$ for gas and dark matter particles respectively. The \textsc{simba} cosmology is flat, with $H_0 = 68 \text{ km/s/Mpc}$ and $\Omega_{m,0}=0.3$.

\subsection{Simulated Galaxies Sample Selection}\label{subsec:simgals}

We identify galaxies within the \textsc{simba} simulation at integer redshifts $z = 0-6$ using \textsc{caesar} \citep{thompson_pygadgetreader_2014}, a package utilizing \textsc{yt} \citep{turk_yt_2011} and a 6D friends-of-friends (FOF) algorithm, to group galaxies and dark matter halos. A `galaxy' is defined as a group of bound particles in the simulations featuring a minimum of 24 star particles. \textsc{caesar} also computes relevant quantities such as the stellar mass and dust mass of the identified galaxies. % check actual number to be sure
In the m25n512 box, \textsc{caesar} identifies 968, 1502, 2065, 2353, 2547, 2924, and 3658 galaxies under this simple definition at $z$ = 6, 5, 4, 3, 2, 1, and 0 respectively.  The minimum stellar mass of a galaxy is then $\sim3\times10^7\text{ M}_{\odot}$ at each redshift.

\begin{figure}
    \epsscale{1.2}
    \plotone{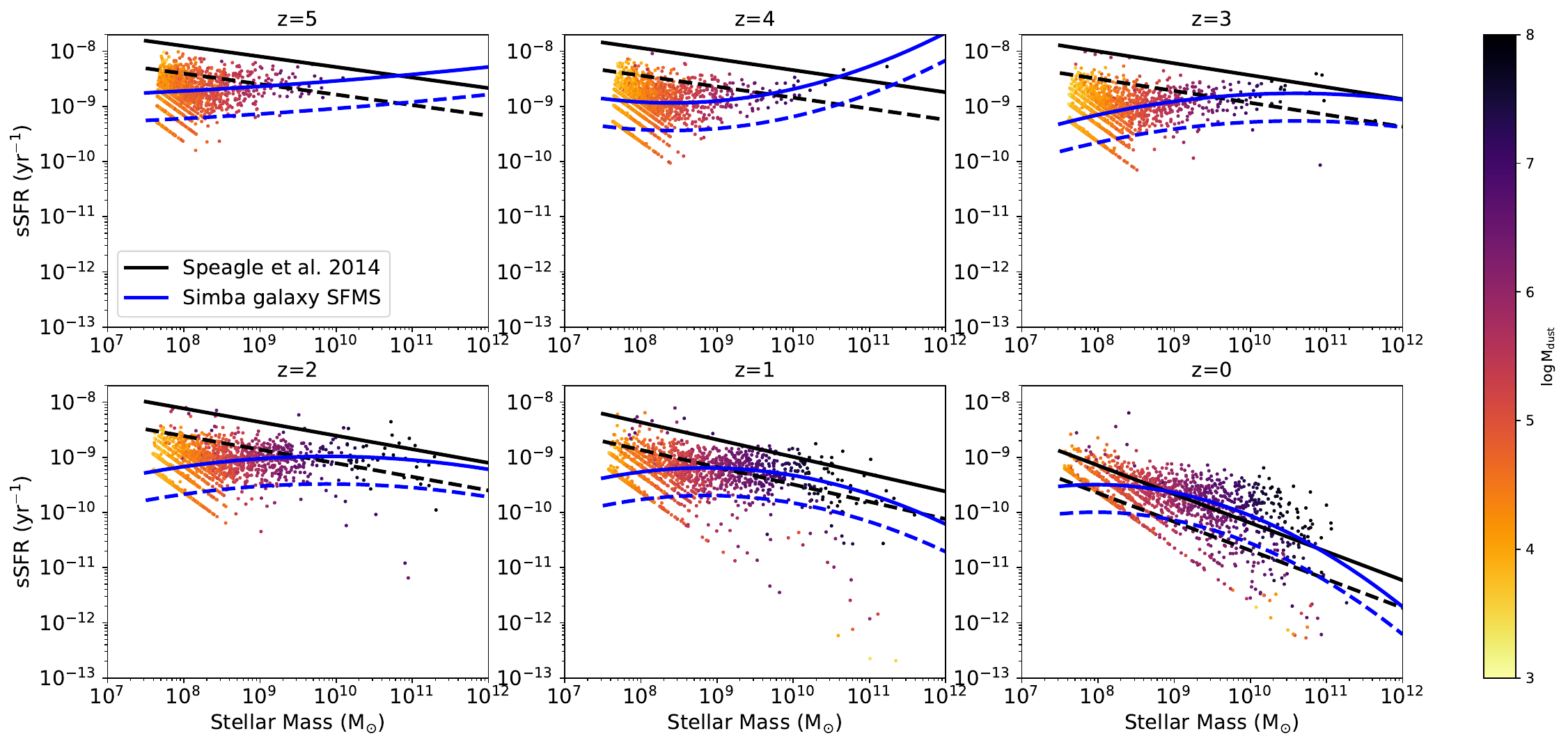}
    \caption{\textbf{Presentation of the derived Simba galaxy star formation main sequence (SFMS).} The blue lines are quadratic fits similar to the work of \cite{whitaker_constraining_2014} and illustrate the \textsc{simba} main sequence and 0.5 dex SF cutoff in solid and dashed lines respectively. Galaxies above the dashed blue line are labeled as star-forming and those below are labeled as quenched. Black lines represent the observationally derived \cite{speagle_highly_2014} SFMS for comparison. The striping visible at low SFR reflects the nature of the simulation forming discrete star particles at a fixed mass resolution. The fit for $z=4$ has an upturn at high masses, but we are focused on a reasonable identification of star-forming galaxies rather than the precise details of the main sequence.}
    \label{fig:simba_ms}
\end{figure}

Most observational analyses of obscured star formation limit themselves to galaxies that are actively forming stars. Galaxies that are actively star-forming will naturally make up the majority of dust-obscured star formation. Additionally, the molecular gas to dust mass ratio between the low and high sSFR galaxies is expected to be very different \citep{whitaker_molecular_2021}. We therefore attempt to limit our analysis to star-forming (SF) galaxies within the simulation. We derive the star formation main sequence of galaxies (SFMS) in the \textsc{simba} simulation as in \cite{akins_quenching_2022}; namely, we iteratively fit a quadratic curve in the shape of \cite{whitaker_constraining_2014}'s Equation 2 at each redshift. We define all galaxies more than 0.5 dex in SFR below the fit for a given stellar mass as `quenched'. Anything above this cutoff we treat as `star-forming' in this work. Notably, this means that this analysis does not exclude starburst galaxies that are experiencing a large uptick in star formation. We present the derived main sequence in Figure \ref{fig:simba_ms}. 

\subsection{Radiative Transfer and SED Processing}\label{subsec:simrad}

The 3D radiative transfer code \textsc{powderday} \citep{narayanan_powderday_2021} enables the generation of synthetic spectral energy distributions (SEDs) for galaxies identified with \textsc{caesar}. The package \textsc{yt} is responsible for interfacing with the simulation output \citep{turk_yt_2011}. \textsc{yt} smooths the discrete particles onto a grid using an octree to prep for the later Monte Carlo radiative transfer step.  The stellar spectra are generated with {\sc fsps},  assuming that each star particle can be treated as a simple stellar population (SSP) \citep{conroy_propagation_2009,conroy_propagation_2010}. For consistency with \textsc{simba}, we use a Chabrier IMF to generate the SSPs \citep{chabrier_imf_2003}. We use the default \textsc{mist} isochrones \citep{paxton_mesa_2011,paxton_mesa2_2013,paxton_mesa3_2015,dotter_mist0_2016,jieun_mist1_2016} and the \textsc{miles} spectral library \citep{vazdekis_miles1_2010,barroso_miles2_2011,vazdekis_miles3_2015}. Given the input stellar SEDs from \textsc{fsps} and the grid from \textsc{yt}, \textsc{hyperion} performs 3D dust radiative transfer by a Monte Carlo algorithm and produces a final SED \citep{robitaille_hyperion_2011}. The Monte Carlo photon processing stops when 99\% of cells have their emission change by less than 1\% between iterations.

We run \textsc{powderday} on every galaxy at integer redshifts from $z=0-6$.  This enables us to construct a catalog of mock galaxy SEDs across different epochs in the \textsc{simba} simulation. We additionally generate these simulated galaxy SEDs at 9 different orientations per galaxy. This allows us to sample different line-of-sight column densities and explore how the fraction of obscured star formation in a galaxy may depend on its orientation. 

\section{Results}\label{sec:res}

\subsection{Evolution of Obscured Star Formation}\label{subsec:obsc}

We first turn our attention to the fraction of dust obscured star formation in our cosmological simulations. To determine this fraction from our simulations, we integrate the generated galaxy SEDs in the appropriate wavelength regimes to determine the rest-frame IR and UV luminosities, which we define as the integrated luminosity between $8-1000$ $\mu$m and $1216-3000 \AA$ respectively. 
Mirroring the work of \cite{whitaker_constraining_2014}, we convert to star formation rates with \cite{kennicutt_global_1998} the $\text{L}_{\text{IR}}$-SFR conversion (equation \ref{eq:sfrir}) and the \cite{bell_sfruv_2005} $\text{L}_{\text{UV}}$-SFR conversion (equation \ref{eq:sfruv}).  We again note that there are many different assumptions that could be made to derive SFR from observations that could introduce systematic differences in results (e.g., \citealt{murphy_sfr_2011}). However, we proceed using those from \cite{whitaker_constraining_2014} for the sake of comparisons to observations.

\begin{equation}\label{eq:sfrir}
    \text{SFR}_{\text{IR}} (\text{M}_{\odot}/\text{yr}) = 1.09\times10^{-10}\times\text{ L}_{\text{IR}} (\text{L}_{\odot})
\end{equation}
\begin{equation}\label{eq:sfruv}
    \text{SFR}_{\text{UV}} (\text{M}_{\odot}/\text{yr}) = 1.09\times10^{-10} \times2.2\text{L}_{\text{UV}} (\text{L}_{\odot})   
\end{equation}
We compute $f_{\text{obs}}$, the `fraction of obscured star formation', via the observationally-motivated definition $\text{SFR}_{\text{IR}}/(\text{SFR}_{\text{IR}}+\text{SFR}_{\text{UV}})$. Using the luminosity-SFR conversions, we can directly calculate $f_{\text{obs}}$ as in equation \ref{eq:f}. 
 \begin{equation}\label{eq:f}
     f_{\text{obs}} = \frac{\text{L}_{\text{IR}}}{\text{L}_{\text{IR}}+2.2\text{L}_{\text{UV}}}
 \end{equation}

\begin{figure}
    \epsscale{1.2}
    \plotone{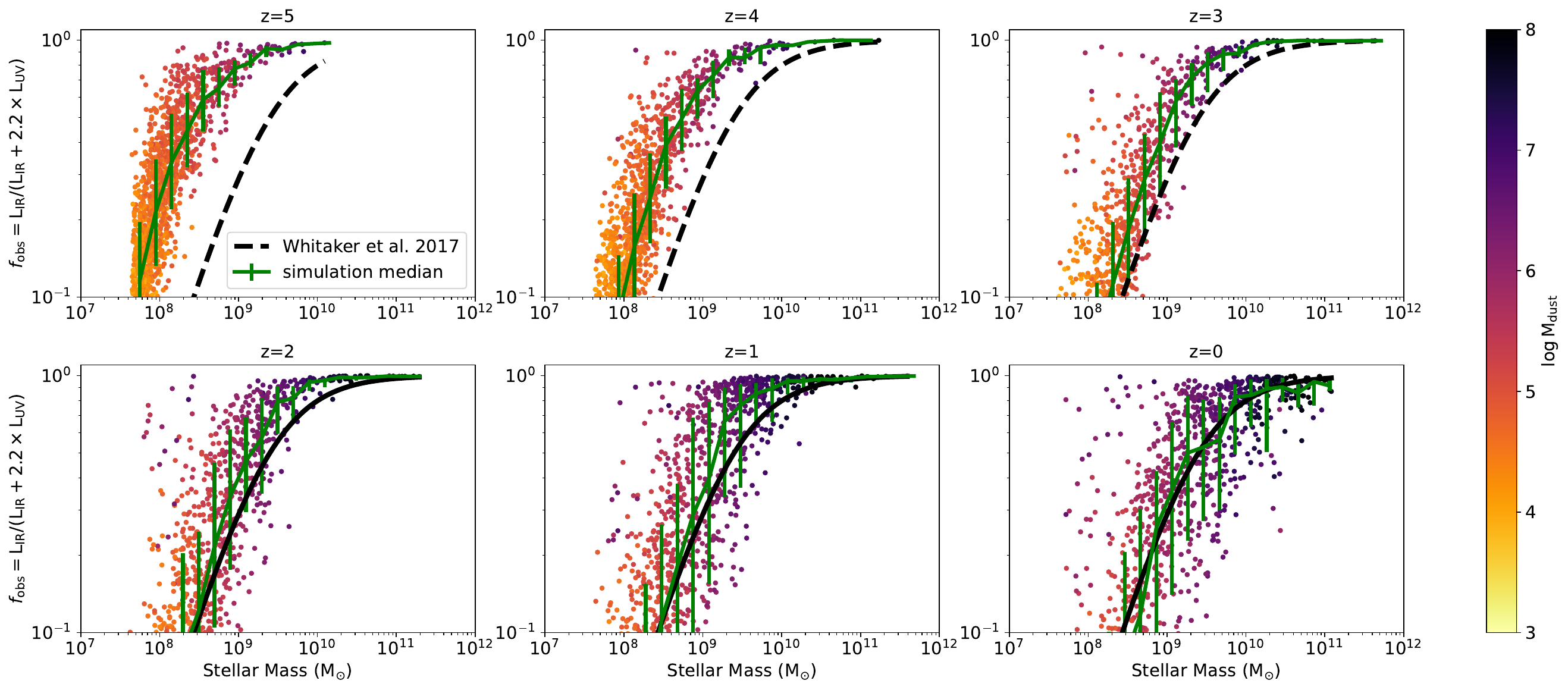}
    \caption{\textbf{$f_{\text{obs}}$, the fraction of obscured star formation in a galaxy, depends on stellar mass and redshift.} Here, the green line indicates a log mass-binned median trend with bins of size 0.2 dex in stellar mass, with error bars marked by the 16th and 84th percentiles of the data within the given bins. The black solid line represents the median trend from observations out to $z=2.5$ as determined by \cite{whitaker_constant_2017}. There is clear evolution in $f_{\text{obs}}$ in galaxies across time; galaxies at fixed stellar mass become less obscured with time. We find the running median for $f_{\text{obs}}$ to reasonably follow the Whitaker fit within the redshift range of the observations ($z=0-2.5)$, though note that our model predicts evolution in $f_{\rm obs}$ at earlier times.
    }
    \label{fig:obs_fig1}
\end{figure}

\begin{figure}
    \epsscale{0.7}
    \plotone{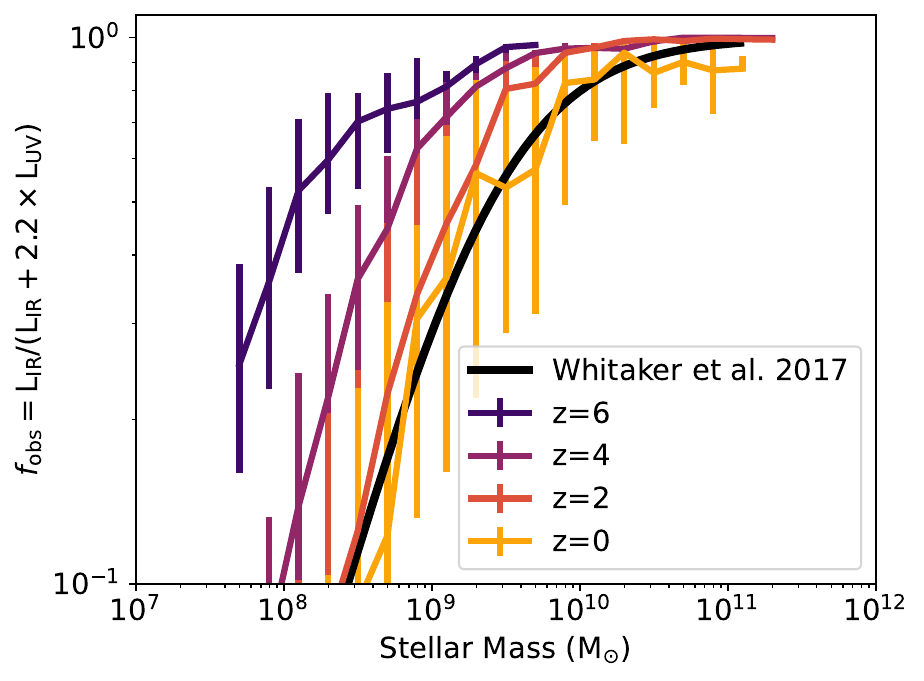}
    \caption{\textbf{Median relationship between $f_{\rm obs}$ and $M_*$ as a function of redshift}. This plot compares $f_{\text{obs}}$ against stellar mass at redshifts $z=0,2,4,6$. At higher redshifts, galaxies at a fixed stellar mass are more obscured than their lower redshift counter parts. These fits are derived by a running median of bin size 0.2 dex in stellar mass, with the error bars produced by the 16th and 84th percentiles of the data belonging to the bin.}
    \label{fig:obs_fig2}
\end{figure}

We now examine our main finding of striking evolution in the fraction of obscured star formation ($f_{\text{obs}}$) at $z>2$. In Figure \ref{fig:obs_fig1}, we compare the stellar mass of our simulated galaxies and their $f_{\text{obs}}$ from $z=0-5$ against the observational median trend derived in \cite{whitaker_constant_2017} from galaxy survey observations out to $z\sim2.5$. It is clear that for our simulation, there is predicted evolution in $f_{\text{obs}}$ with time for galaxies of fixed stellar mass; namely, galaxies at a fixed stellar mass are predicted to have more of their star formation obscured by dust at higher redshifts as compared to lower-redshift galaxies. This redshift evolution is more pronounced at high redshifts, with the offset in the median relation from best fit of \citet{whitaker_constant_2017} only becoming pronounced at $z>2$. The limited evolution in $f_{\text{obs}}$ from $z=0-2$ allows the median simulation trend to be a reasonably consistent with the observed fits from \citet{whitaker_constant_2017} at $z=0-2$, but the highest-$z$ galaxies clearly do not fall on the median observational trend. Note that the best-fit observed trend is indicated with a dotted line for $z=3-5$ owing to the fact that \cite{whitaker_constant_2017} only considered data out to $z=2.5$. 
This point is amplified in Figure \ref{fig:obs_fig2}, where we show the median trends in the $f_{\rm obs}-\text{M}_*$ relation for our simulated galaxies at $z=0,2,4,6$. The $z=0$ and $z=2$ trends roughly follow the observational results of \citet{whitaker_constant_2017}, but $z=4$ and $z=6$ are significantly different from observational results at low redshift; we note the differences of the $z=6$ results being particularly pronounced. The disagreement becomes worse when considering high redshift results, which seem to indicate $f_{\text{obs}}$ increasing at high redshift.

We predict an increase in the dispersion of the $f_{\text{obs}}-\text{M}_*$ relation at lower redshifts. We demonstrate this by comparing the 16th and 84th percentiles of the bins at $z=0$ and $z=5$ in Figure \ref{fig:obs_fig1}. In Figure \ref{fig:obs_fig2}, the addition of the $z=6$ results further supports this result; the $z=0$ dispersion is larger than the $z=6$ as well as the extent of both the other redshifts depicted. 

\subsection{Explaining the Evolution of $f_{\text{obs}}$}\label{subsec:evo_ex}
 We have seen that our cosmological simulation exhibits an evolution in the fraction of obscured star formation; galaxies of the same stellar mass have greater fractions of their UV light obscured by dust at higher redshift. With our evolving, self-consistent dust model, we investigate the physical reason for this. At the outset, there are two likely reasons for this evolution: an evolution in the total dust mass at a fixed stellar mass, and/or an evolution in the dust attenuation law (which manifests evolution in the star-dust geometry \citep{salim_dust_2020}). We now investigate these possible physical origins of the evolution of the $f_{\text{obscured}}-\text{M}_*$ relationship with redshift in turn.

\subsubsection{Evolution of Dust Mass}\label{subsubsec:dustquan}
\begin{figure}
    \epsscale{0.7}
    \plotone{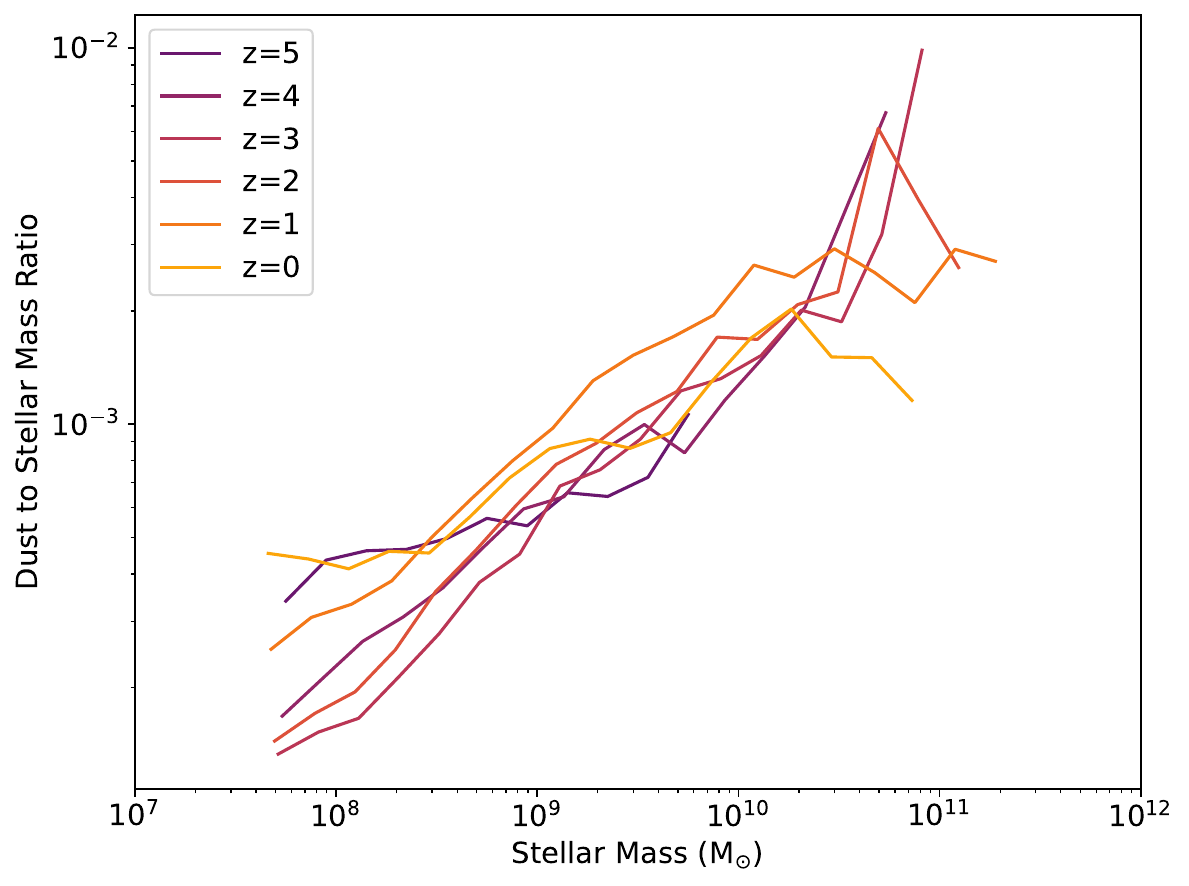}
    \caption{\textbf{The dust mass explains the dependence of $f_{\text{obs}}$ on stellar mass but not the redshift evolution of $f_{\text{obs}}$ at fixed stellar mass.} Here, we compare the stellar mass and dust to stellar mass ratio at various redshifts. Higher mass galaxies have higher fractions of dust than lower mass galaxies. The greater dust fractions also generally correspond with higher $f_{\text{obs}}$. However, for fixed stellar mass of higher mass galaxies, the dust fraction does not evolve with redshift.}
    \label{fig:smdm_rat}
\end{figure}

One possible explanation for the evolution of $f_{\rm obs}$ with redshift at a fixed stellar mass is via an evolution in galaxy dust masses. Larger quantities of dust would correspond to higher typical column densities along the line of sight and therefore more attenuated starlight in the UV. We find an unsurprising correlation between dust-obscured star formation fraction and dust mass. In Figure \ref{fig:obs_fig1}, we color-code the individual galaxies by their dust mass. Most of the highly obscured galaxies have the highest amount of dust accumulated within them. This is especially notable at lower redshifts, where the intrinsic scatter is larger; many of the simulated galaxies that are outliers to the median trend in obscuration have noticeably higher dust masses than their less obscured counterparts at similar stellar mass. This result is reinforced with Figure \ref{fig:smdm_rat}, where we analyze the dust-to-stellar mass ratio across redshift and find a strong correlation between stellar mass and dust mass to stellar mass ratio. This correlation can allow us to physically explain the trend of $f_{\text{obs}}$ with stellar mass; more massive galaxies in our simulations have more dust per stellar mass, which naturally results in higher $f_{\text{obs}}$. However, this does not explain the evolution in $f_{\text{obs}}$ with redshift. Figure \ref{fig:smdm_rat} exhibits no clear trend with redshift at a fixed stellar mass. Similarly, \cite{li_dust--gas_2019} find that the dust-to-gas mass ratio in the \textsc{simba} simulation does not vary strongly with redshift. This rules out a change in dust mass with time as an explanation for the strong evolution of the median $f_{\text{obs}}$ trend at $z>2$. 

\subsubsection{Star-Dust Geometry}\label{subsubsec:sdgeo}
We now turn to evolution in the star-dust geometry as an explanation for the redshift evolution of the $f_{\mathrm{obs}}$-M$_{\star}$ relation. The general premise here is that two galaxies with the same $\text{M}_{\rm dust}$ but different dust attenuation laws would result in different ratios of IR to UV flux.  Star-dust geometry can manifest itself in two manners: galaxy viewing orientation, and ISM clumping.  We explore these in turn.

\begin{figure}
    \epsscale{1.2}
    \plotone{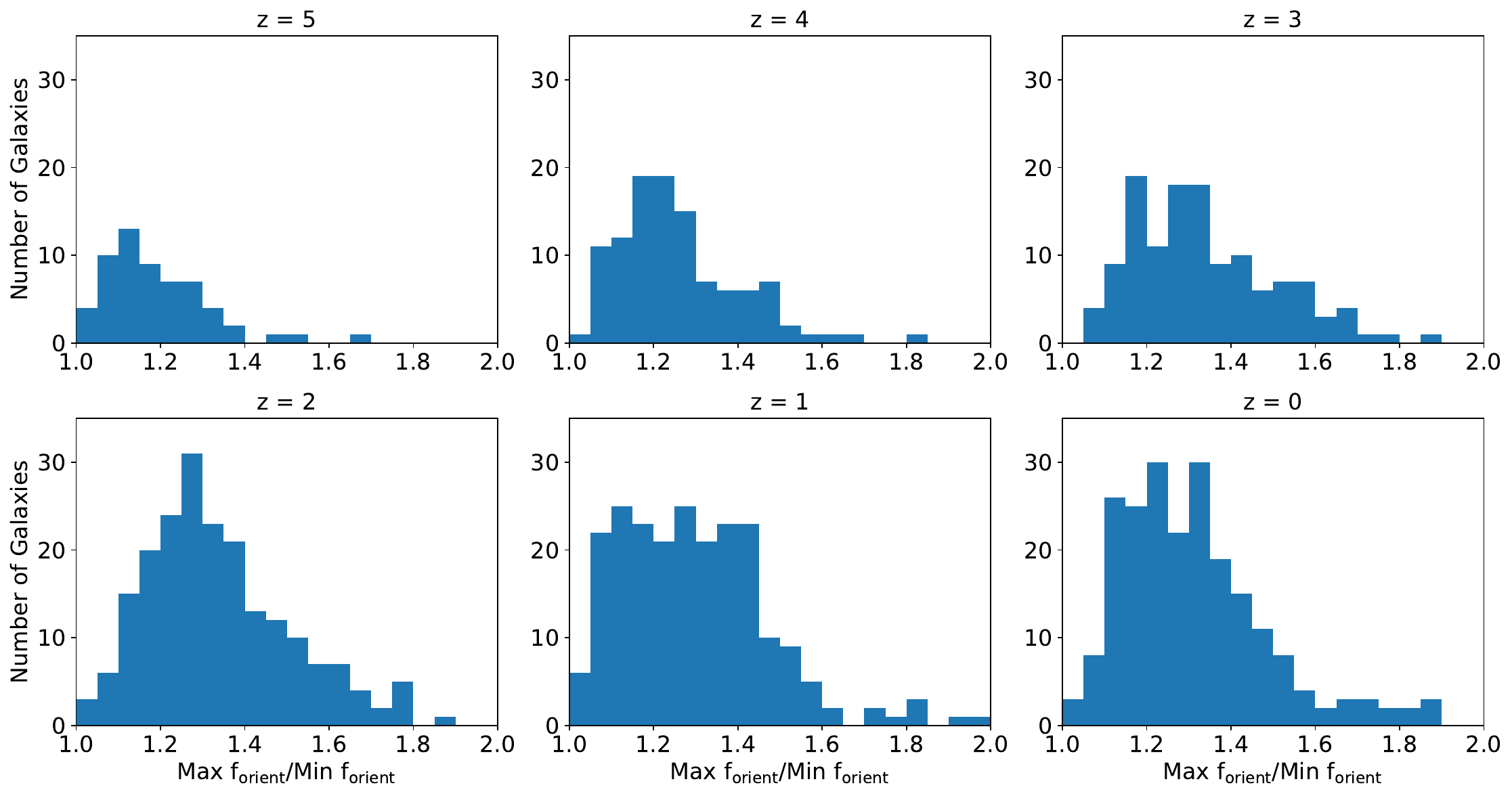}
    \caption{\textbf{Orientation effects cannot explain the trends in $f_{\text{obs}}$ with redshift.} We plot for galaxies in the range $8.75\le \log{\text{M}_{*}}\le9.25$ the ratio between the maximum derived $f_{\text{obs}}$ and the minimum derived $f_{\text{obs}}$ of the 9 orientations of the system we run \textsc{powderday} on. High values of this ratio indicate a high dependence of the $f_{\text{obs}}$ of the galaxy on the viewing angle of the observer.}
    \label{fig:orient_m9}
\end{figure}

We first explore the effects of orientation on the fraction of obscured star formation. Star-dust geometry would be expected to vary depending on the viewing angle of the galaxy. If high-redshift galaxies in our simulations tend to be less disky, then they would have less of a viewing-angle bias in their $f_{\text{obs}}$. Conversely, if the fraction of galaxy disks is higher at a given redshift, this would result in a greater spread based on viewing angle and more viewing angles with a lower column density of dust. To test this, we have set up cameras at $9$ isotropically arranged viewing angles around each galaxy. In Figure \ref{fig:orient_m9}, we present the ratios between the maximum and minimum $f_{\text{obs}}$ among these $9$ orientations for galaxies whose stellar mass is $\sim10^9\text{ M}_\odot$. A value of 1 means there is no dependence on the viewing angle for $f_{\text{obs}}$; the greater the value, the more important the viewing angle of the galaxy is to the observed SED. A factor of 2 indicates that $f_{\text{obs}}$ is doubled by orienting the galaxy a different way. Though it is clear that viewing angle can have a significant impact on $f_{\text{obs}}$ in any individual galaxy, there is no evolution visible in this bin of stellar mass across redshift. Therefore, we conclude that orientation is not the reason for the evolution of $f_{\text{obs}}$. However, it is likely that some portion of the spread in the median $f_{\text{obs}}$ relation can be explained by viewing angle.

\begin{figure}
    \epsscale{1.2}
    \plotone{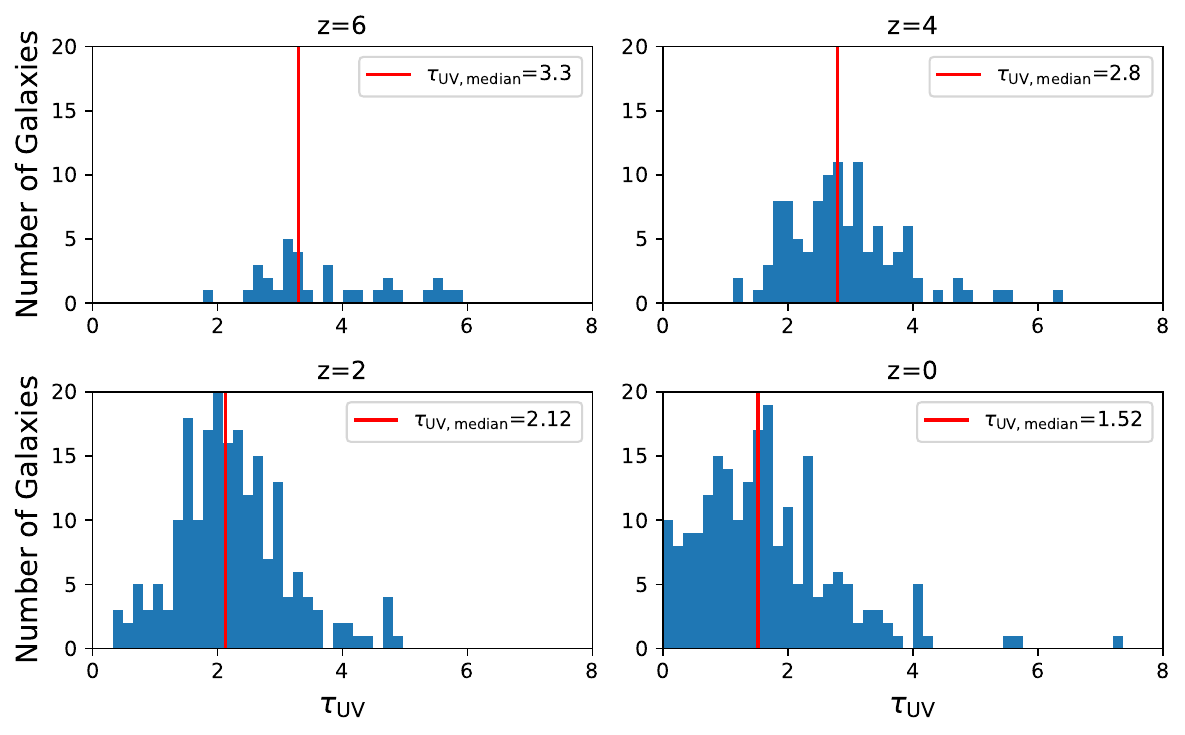}
    \caption{\textbf{The evolution in UV optical depths explains the evolution of $f_{\text{obs}}$ via evolution in star-dust geometries.} We construct a histogram of galaxies' optical depth in the UV range around $\lambda=1500 \text{ }\mu \text{m}$ for galaxies in the range  $8.75\le \log{\text{M}_{*}}\le9.25$. There is a clear shift in the galaxies toward lower UV attenuation with lower redshift, which reflects the evolution in obscuration.}
    \label{fig:tau}
\end{figure}

We now examine turn to the impact of ISM clumping, which we quantify via the UV optical depth. With our knowledge of the intrinsic stellar SEDs before dust radiative transfer, we are able to directly compute the attenuation curves of our simulated galaxies. The optical depths in this paper are calculated by the standard equation \ref{eq:tau}, where $I_{\lambda,0}$ represents the intrinsic stellar SED from the total of the SSP spectra that make up the galaxy produced by \textsc{fsps}, and $I_{\lambda}$ represents the final SED after radiative transfer is performed.
\begin{equation}\label{eq:tau}
    \tau_\lambda = -\ln{\frac{I_\lambda}{I_{\lambda,0}}}
\end{equation}
In Figure \ref{fig:tau}, we analyze the optical depths of the galaxies in the UV regime in a stellar mass bin around $10^9 \text{ M}_{\odot}$. We do see a clear shift in the distribution with redshift. The evolving optical depths indicate that the column density of dust along the line of sight to young stars is, on average, decreasing with redshift for galaxies of a fixed stellar mass. This change in column density, taken with the lack of evolution in the dust-to-stellar mass ratios, highlights the role of star-dust geometry driving the increased obscuration at high-$z$. We therefore conclude that the star-dust geometry is playing the main role in our evolution of the obscuration fraction with redshift by the evolution of the dust column densities.

\begin{figure}
    \epsscale{0.8}
    \plotone{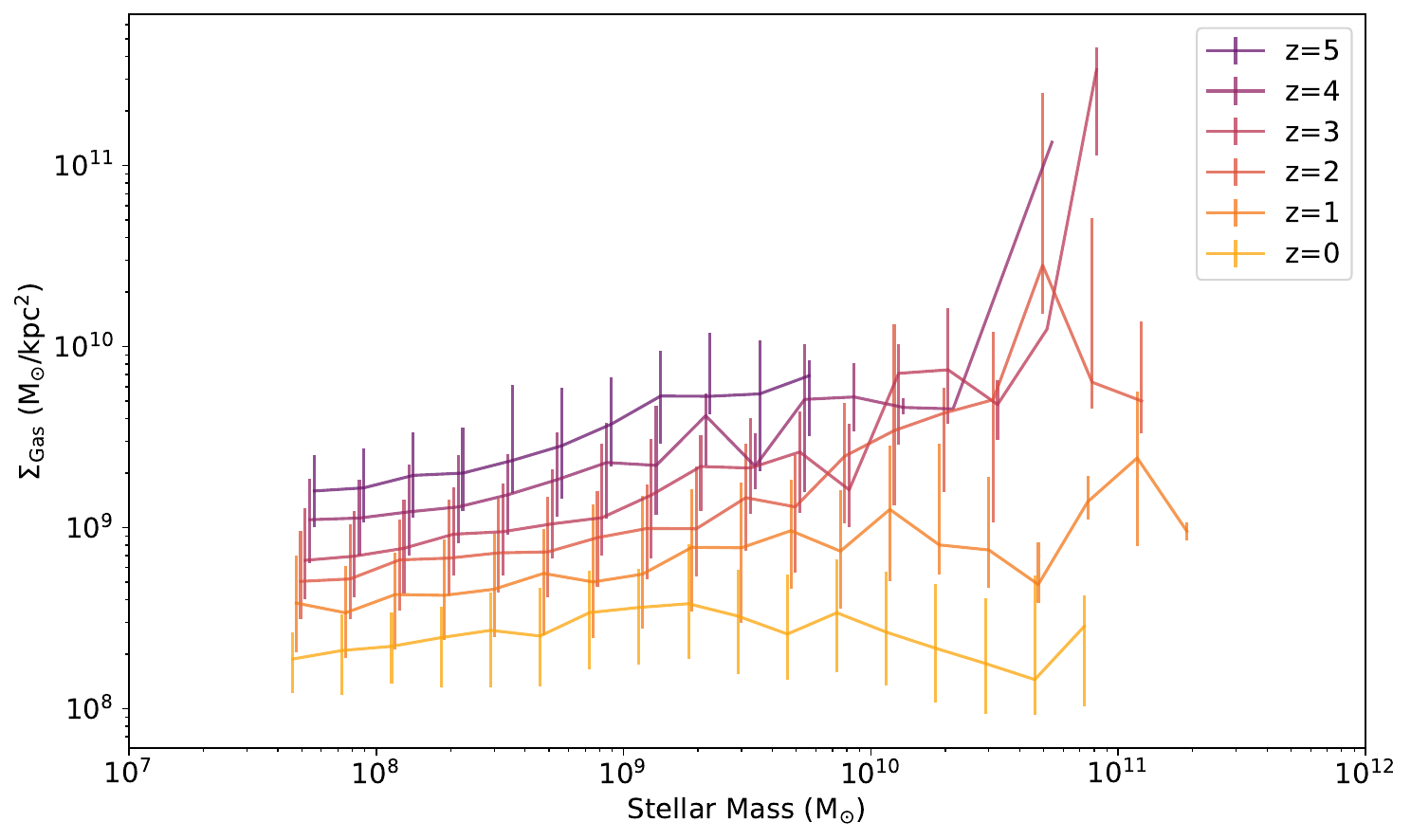}
    \caption{\textbf{The evolution in the gas surface density corresponds to the evolution in $f_{\text{obs}}$}. We compute the galactic median trend in surface density across redshifts. We estimate a gas surface density by dividing the gas mass by the square of the radius that encloses on half the gas mass. The error bars represent the 16th and 84th percentiles for the galaxies that fall in the bin. There is a monotonic trend in the gas surface density with time, which follows the explanation presented by \cite{shapley_mosfire_2022}.}
    \label{fig:gas_surf}
\end{figure}

This result is also consistent with the model presented by \cite{shapley_mosfire_2022} that $ \tau \propto \kappa_{\lambda} (\text{M}_{\text{dust}}/\text{M}_{\text{gas}}) \Sigma_{\text{Gas}}$. In our work, the extinction is constant in each cell, and we have already shown that the dust to stellar mass ratio exhibits no evolution; this would assign all evolution in attenuation to the gas surface density. In Figure \ref{fig:gas_surf}, we compute the gas surface density as the ratio of the gas mass and the square of the gas half-mass radius.  Indeed, we note that the gas surface density of the simulated galaxies is decreasing with redshift at a fixed stellar mass.

\subsection{Obscured Cosmic Star Formation Rate Density}\label{subsec:sfr_res}

\begin{figure}
    \plotone{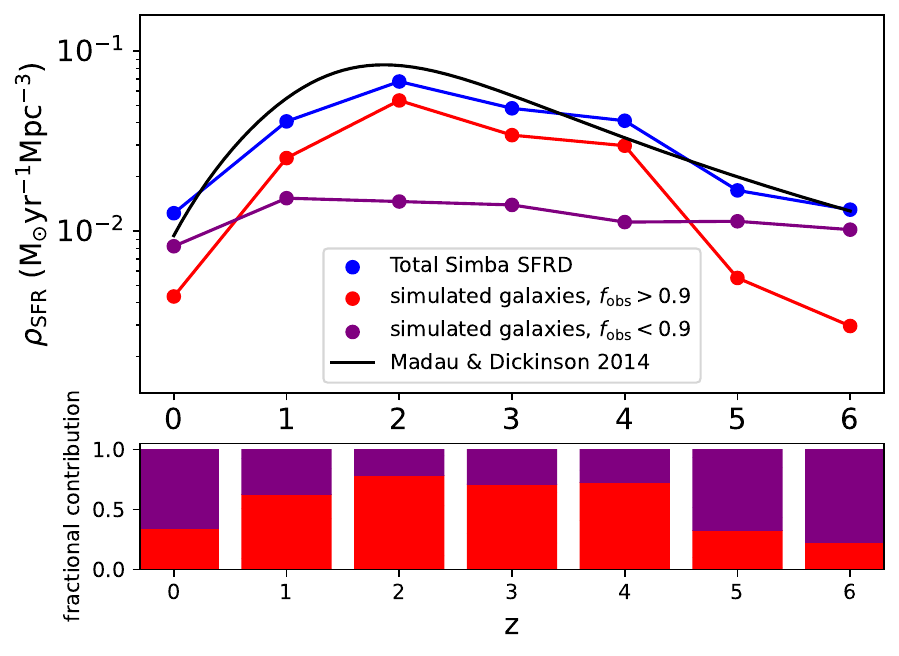}
    \caption{\textbf{Most obscured star formation dominates the SFR budget of the universe below $z=4$.} We plot the SFRD across cosmic time in the \textsc{simba} simulations. We break up the galaxies' SFR averaged over 100 Myr into the population of highly obscured galaxies and others. After $z\sim4$, the largely obscured galaxies dominate the total SFR density down to $z\sim0$. The behavior of the total SFRD roughly traces the fit from \cite{madau_cosmic_2014}. The \citet{madau_cosmic_2014} SFR trend is scaled to a Chabrier IMF by the conversion factor 0.63 as described in the \cite{madau_cosmic_2014}.
    }
    \label{fig:madau}
\end{figure}

We combine our results for individual galaxies to investigate the contribution of total obscured star formation to the total cosmic star formation rate density.  We separately sum and plot the SFR derived from our \textsc{caesar} group finder for all highly obscured galaxies ($f_{\text{obs}}>90\%$) and the remaining galaxies at each redshift.  To convert to a star formation rate density (SFRD), we divide these results by the co-moving volume of the \textsc{simba} m25n512 run. The results are displayed in Figure \ref{fig:madau}. At $z=4$, there is a turnover where galaxies with obscured star formation become the main contributors to the cosmic star formation rate density (CSFRD) in the simulation. This qualitatively matches with observational results \citep{dunlop_deep_2017,zavala_evolution_2021}. Additionally, the \textsc{simba} cosmic SFRD appears to reasonably trace the observationally derived SFRD reasonably well out to $z\sim6$. Importantly, obscured star formation dominates the SFRD budget in the epoch of `cosmic noon'; this suggests that most of the star formation in the history of the universe is obscured. We attribute this turnover to the relative lack of galaxies sufficiently massive to match our criteria of $f>0.9$ as the designation of an `obscured galaxy' despite the higher obscuration at fixed stellar mass at high redshift.

\section{Discussion}\label{sec:dis}

\subsection{Comparisons to Obscured Star Formation Observations}\label{subsec:obsdis}
In this section, we place our results for the evolution of the obscured fraction of star formation, $f_{\text{obs}}$, in observational context.
We conclude that there is evolution in the $f_{\text{obs}}$ across cosmic time, though the change does not appear significantly different from previous results up to $z\sim3$.

Our results are reasonably consistent with observations of $f_{\text{obs}}$ at lower redshift. \cite{whitaker_constant_2017} studied surveys of SF galaxies selected by color up to redshift $z\sim2.5$ using the same $\text{L}_
{\text{UV}}$-SFR  and $\text{L}_{\text{IR}}$-SFR conversions referenced in this paper. \citet{whitaker_constant_2017} find no significant evolution in $f_{\text{obs}}$ in this redshift range; similarly, our results show only mild evolution of $f_{\text{obs}}$ in this range. Therefore, our results could be seen as consistent with the Whitaker observations at these lower redshifts and a prediction for greater $f_{\text{obs}}$ for galaxies that exist in the first few billion years. For further comparison, we fit our model results of $f_{\text{obs}}-\text{M}_*$ with the form of equation \ref{eq:whitfit} and present these best-fit parameters in Table~\ref{tab:fit_params}.
\begin{equation}
    \label{eq:whitfit}
    f_{\text{obs}} = \frac{1}{1+ae^{b\log{\text{M}_*/M_{\odot}}}}
\end{equation}

\begin{table}
    \centering
    \begin{tabular}{|c|c|c|p{0.45\linewidth}}
        \toprule
        {\bf z} & {\bf a} & {\bf b}\\
        \midrule
        6 & $8.772\times10^{7}$ & $-2.253$ \\\midrule
        5 & $3.626\times10^{9}$ & $-2.621$ \\\midrule
        4 & $1.017\times10^{10}$ & $-2.642$ \\\midrule
        3 & $1.860\times10^{10}$ & $-2.621$ \\\midrule
        2 & $8.690\times10^{9}$ & $-2.511$ \\\midrule
        1 & $1.052\times10^{10}$ & $-2.528$ \\\midrule
        0 & $7.432\times10^{7}$ & $-1.932$ \\\midrule
        \cite{whitaker_constant_2017} & $1.96\times10^9$ & $-2.277$ \\\midrule

    \end{tabular}
    \caption{Best fit parameters for equation \ref{eq:whitfit} for the obscured fraction of star formation as a function the galactic stellar mass across different redshifts from simulation results. We include the results from \cite{whitaker_constant_2017} for comparison.}
    \label{tab:fit_params}
\end{table}

\begin{figure}
    \epsscale{1.2}
    \plotone{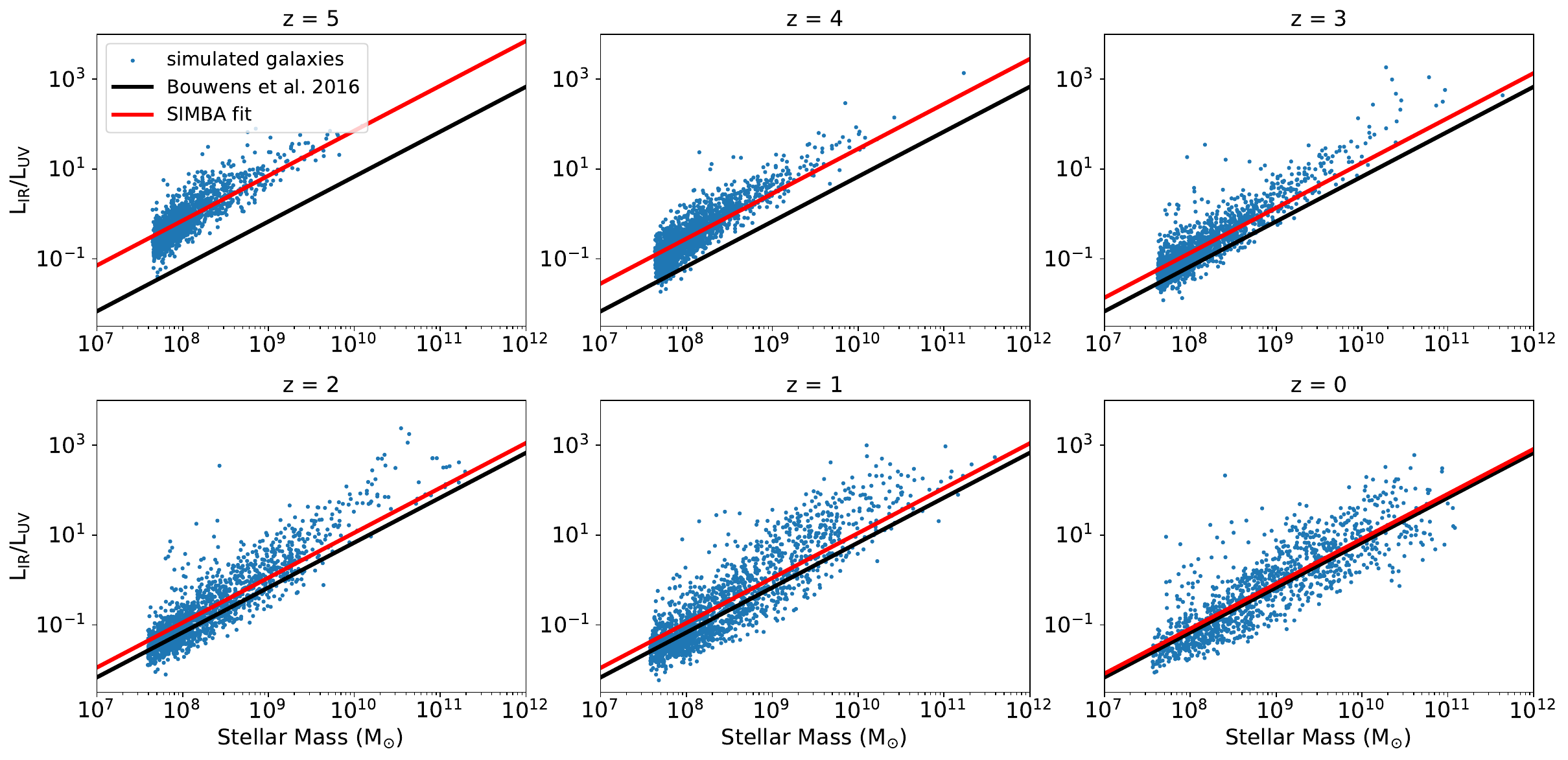}
    \caption{\textbf{Other expressions of obscured star formation are also inconsistent with the notion of constant obscuration at higher redshift}. IRX-$\text{M}_*$ relation plotted against z and compared to the \cite{bouwens_alma_2016} fit derived from a collection of $z\sim2-3$ galaxies. Similarly to the fraction of obscured star formation, the best fit at each redshift to equation \ref{eq:irx_bouw} is reasonably consistent with this fit out to $z\sim3$, and differs noticeably at higher redshifts.}
    \label{fig:irxm}
\end{figure}

Our results are similarly comparable to other studies for low-redshift obscured star formation. \cite{shapley_mosfire_2022} explore the lack of evolution through attenuation curves and compare the SDSS $z\sim0$ to the MOSDEF $z\sim2.3$ sample, and also find no significant evolution in obscured star formation. 
A similar parameter to $f_{\text{obs}}$ that is derived from galactic observations is the `infrared excess' (IRX), defined in equation \ref{eq:irx}. \cite{bouwens_alma_2016} present a best fit to the infrared excess at all times in the form of equation \ref{eq:irx_bouw} combining several studies of galaxies from $z\sim2-3$. \cite{bouwens_alma_2016} also concluded that the relation between stellar mass and the observed infrared excess does not evolve significantly over cosmological time if they assume dust temperature evolution with redshift. Given that the infrared excess is comprised of the same quantities as the $f_{\text{obs}}$, we also see evolution in the IRX - $\text{M}_*$ relation in the \textsc{simba} simulations (see Figure \ref{fig:irxm}). Our results again only start to show strong divergence from the \citet{bouwens_alma_2016} trend at higher redshift.
\begin{equation}
    \label{eq:irx}
    \text{IRX}=\frac{\text{L}_{\text{IR}}}{\text{L}_{\text{UV}}}
\end{equation}

\begin{equation}
    \label{eq:irx_bouw}
    \log_{10}{\text{IRX}}=\log_{10}{\frac{\text{M}_*}{\text{M}_{\odot}}} - \text{C}
\end{equation}

Contrary to our results, observational studies of higher redshift galaxies tend to find $f_{\text{obs}}$ to be lower than the previously discussed low-z observations, which is in direct tension with our findings. The results from the ALPINE survey at higher redshift suggest that the typical $f_{\text{obs}}$ of galaxies at fixed stellar mass is actually lower \citep{fudamoto_alpine_2020,khusanova_obscured_2021}.  This may owe to observational selection effects. For example, the ALPINE survey was based off of a set of rest-UV selected galaxies; this inherently would bias the resultant sample to galaxies with lower $f_{\text{obs}}$. \cite{pope_early_2017,pope_obs_2023} find a $z=4$ lensed galaxy with similarly high $f_{\text{obs}}$ to our results.  \cite{williams_missed_2023} suggest that ALMA and Hubble do not observe a significant population of obscured galaxies. Observations by JWST should provide updated estimates for the contributions of obscured galaxies and will serve as a test for our predictions.

\cite{pallottini_serra_2022} modeled zoom simulations of high-redshift galaxies, which includes many obscured galaxies with high star formation rates. They assign dust to their simulated galaxy by assuming a dust-to-metals ratio. From spectra of their galaxies, \citet{pallottini_serra_2022} compute the IRX-$\beta$ relation and find that much of the star formation is heavily obscured by dust despite on average relatively low attenuation. The range in IRX and stellar mass in their sample is relatively analogous to our results. These results further underscore the importance of star-dust geometry to the overall results. However, we find that our obscured galaxies do typically have high average optical depths, so the low attenuation discussed in \citet{pallottini_serra_2022} is in tension with our results. The likely main causes of this strong tension are the details of the dust model and the difference in resolution between the cosmological \textsc{simba} simulation and the zoom-in SERRA simulation.

\subsection{Cosmic Star Formation Rate Density vs Observations}\label{subsec:csfrd}
We now turn to comparing our results for the obscured CSFRD against observational results.
The star formation rate density of the \textsc{simba} simulation m25n512 box reasonably traces the observational results, both in terms of qualitative behavior over cosmic time of the SFRD and the turnover between the dominance of obscured and unobscured star formation at $z\sim4$ \citep{dunlop_deep_2017,zavala_evolution_2021}.

\cite{casey_bright_2018} discuss two potential extreme models for the buildup of dusty galaxies in the early universe: a `dust-poor' early universe and a `dust-rich' early universe. The dust-poor model involves dusty galaxies only becoming the major contributor to star formation at cosmic noon; the dust-rich model instead involves dusty starbursts playing the main role at high redshifts while dusty galaxies would be the main contributors in the range of $1.5<z<6.5$. Our predictions include aspects from both of these models, though they favor a universe more closely resembling the dust-rich model. At all redshifts in this work, the 100 galaxies with the greatest star formation contribute the majority of star formation in the simulation (like the dust-rich model). However, it is around cosmic noon where the the total star formation can largely be attributed to the galaxies with the most active star formation (like the dust-poor model).  Additionally, we predict that obscured star formation is dominant starting at $z=4$ only despite the fact that our high-z galaxies are highly obscured rather than UV-bright. These differences in our results likely could be used to inform constraints on the parameter space in the \cite{casey_bright_2018} model.

\cite{dunlop_deep_2017} use deep ALMA data corresponding to the Hubble Ultra Deep Field to identify a total of 16 galaxies in the 1.3 mm range. All galaxies they identify have estimated stellar masses $\gtrsim10^{9.6}\text{ M}_{\odot}$. Using these extreme galaxies, they identify a turnover at $z\sim4$ where dust-obscured star formation contributes to the majority of the total star formation. They offer the explanation that the high contribution to obscured galaxies is not dependent on the total dust mass, but instead on the fast growth of higher-mass galaxies that contain most of the star formation around and before cosmic noon. Our results generally agree with this notion; as discussed, we do not find a strong dependence of the dust mass with redshift at fixed stellar mass, and we see several massive galaxies around the epoch of cosmic noon with incredibly high star formation rates that contribute significantly to the total star formation rate.

High-redshift individual galaxy detections such as \cite{fudamoto_normal_2021} have detected early-time, dusty, star-forming galaxies. \cite{fudamoto_normal_2021} estimate that obscured star formation at $z > 6$ could be contributing between $10\%$ and $25\%$ to the total density. However, \cite{gruppioni_alpine-alma_2020} in the ALMA ALPINE survey identify higher fractions of obscured star formation using a larger sample of $\sim50$ detected infrared sources. Our results agree better with the former conclusion than the latter explanation, though they are sensitive to the our conversion of simulated results to total obscured and unobscured star formation.

Several theoretical works have presented predictions for the contribution of obscured star formation to the total star formation budget at high redshift. \cite{shen_jwsttng_2022} study obscured star formation via forward modeling the IllustrisTNG simulation assuming a constant dust to metals ratio that evolves with redshift at $z=4,6,8$. They find that they underpredict the abundance of the most IR-luminous galaxies. Using the \cite{murphy_sfr_2011} conversion, \citet{shen_jwsttng_2022} find roughly equal contributions of unobscured and obscured star formation at $z\sim4$ (estimated based on their Figure 8) and predict unobscured star formation to be be dominant for $z\gtrsim4$. This is in keeping with our results; both works lack a sample of high-redshift high $\text{L}_{\text{IR}}$ galaxies and find obscured star formation to be subdominant at very high redshift.

Other theoretical predictions for obscured star formation at individual high redshifts have varying levels of agreement with our results. \cite{mauerhofer_delphi_2023} build a semi-analytic model (SAM) for galaxy and dust evolution and generate UV and IR luminosities by assuming a single dust grain type and size and a dust geometry that produces a single UV photon escape fraction. \citet{mauerhofer_delphi_2023} predict $34\%$ of star formation to be obscured at $z=5$, which is consistent with our results. \citet{ma_fire2lum_2018} present the FIRE-2 zoom simulation results. Their spectra are generated by a combination of stellar population synthesis models assumption of a dust-to-metals ratio of 0.4 for Small Magellinic Cloud-like dust. \citet{ma_fire2lum_2018} find $37\%$ of UV light from bright galaxies to be obscured at $z=6$. This is noticeably higher than the results we discuss in this paper.
\citet{lewis_dustier_2023} run a simulation with a dust production/destruction model included and generate spectra for their halos by taking the intrinsic spectra and computing the extinction along the line of sight for each halo. \citet{lewis_dustier_2023} calculate the overall fraction of obscured star formation and predict $\sim45\%$ to be obscured at $z=5$ and $\sim40\%$ at $z=6$ from magnitude-cut sample of galaxies; our results are slightly lower than these predictions. Our dust model's differences from the dust models of these theoretical works likely accounts for many of the differences in our results.

\section{Summary}\label{sec:sum}
We utilized the galaxies from the \textsc{simba} simulation in integer redshifts from $z=0$ to $z=6$ coupled with synthetic \textsc{powderday} SEDs to study obscured star formation throughout cosmic time. 

\begin{itemize}
    \item We find significant evolution in the fraction of obscured star formation at $z>2$. Star-forming galaxies at fixed stellar mass are more obscured at higher redshifts, and the spread in $f_{\text{obs}}$ along the median trend decreases with increased redshift (\S~\ref{subsec:obsc}, Figure \ref{fig:obs_fig1}).
    \item We explain the observed trend between $f_{\text{obs}}$ and stellar mass at fixed redshift by the increasing dust fractions at higher galactic stellar mass  (\S~\ref{subsubsec:dustquan}, Figure \ref{fig:smdm_rat}). This behavior is present at all redshifts we explore.
    \item As the dust mass does not strongly evolve at fixed stellar mass with redshift, we explain the evolution in $f_{\text{obs}}$ through evolving star-dust geometry. The column density of dust along the line of sight to young stars is decreasing on average with time (\S~\ref{subsubsec:sdgeo}, Figure \ref{fig:tau}).
    \item Obscured star formation is the dominant contributor across the epoch of cosmic noon to the total star formation budget, with a turnover corresponding to $z\sim4$, similar to observational results (\S~\ref{subsec:sfr_res}, Figure \ref{fig:madau}).
    \item The evolution in the $f_{\text{obs}}$ is weak enough at lower redshifts to explain why previous observational results studying lower redshifts ($z\lesssim3$) find no significant evolution in obscuration over most of cosmic time (\S~\ref{subsec:obsdis}).
\end{itemize}

\section{Acknowledgements}
This work is funded by NSF AST-1909141.   DN
additionally thanks the Aspen Center for Physics which is supported by
National Science Foundation grant PHY-1607611, which is where the
original framework for the {\sc powderday} code base was developed.

\bibliography{export_data.bib}

\end{document}